\documentclass[prd,aps,nofootinbib,showpacs,epsf,floats,amsfonts,amssymb,amsmath]{revtex4}

\usepackage{epsfig}

\newcommand{\be}{\begin{equation}}
\newcommand{\ee}{\end{equation}}
\newcommand{\bea}{\begin{eqnarray}}
\newcommand{\eea}{\end{eqnarray}}
\newcommand{\beaa}{\begin{eqnarray}}
\newcommand{\eeaa}{\end{eqnarray}}
\newcommand{\ba}{\begin{array}}
\newcommand{\ea}{\end{array}}
\newcommand{\bit}{\begin{itemize}}
\newcommand{\eit}{\end{itemize}}
\newcommand{\ben}{\begin{enumerate}}
\newcommand{\een}{\end{enumerate}}

\def\lab{\label}
\def\lan{\langle}

\def\Lrar{\Leftrightarrow}

\def\non{\nonumber}

\def\pa{\partial}
\def\ran{\rangle}
\def\rar{\rightarrow}

\def\ti{\tilde}

\def\al{\alpha}

\def\de{\delta}
\def\De{\Delta}
\def\ep{\epsilon}

\def\te{\theta}

\def\la{\lambda}

\def\vec#1{{\bf #1}}

\begin{document}


\title{Topological defects, fractals and\\ the structure of quantum field
theory}

\author{Giuseppe Vitiello
\vspace{3mm}}

\address{Dipartimento di Matematica e Informatica\\Istituto Nazionale di
Fisica Nucleare, Gruppo Collegato di Salerno
\\Universit\'a di
Salerno, 84100 Salerno, Italia\\ vitiello@sa.infn.it -
http://www.sa.infn.it/giuseppe.vitiello}




\begin{abstract}
In this paper I discuss the formation of topological defects in
quantum field theory and the relation between fractals and coherent
states. The study of defect formation is particularly useful in the
understanding of the same mathematical structure of quantum field
theory with particular reference to the processes of non-equilibrium
symmetry breaking. The functional realization of fractals in terms
of the $q$-deformed algebra of coherent states is also presented.
From one side, this sheds some light on the dynamical formation of
fractals. From the other side, it also exhibits the fractal nature
of coherent states, thus opening new perspectives in the analysis of
those phenomena where coherent states play a relevant role. The
global nature of fractals appears to emerge from local deformation
processes and fractal properties are incorporated in the framework
of the theory of entire analytical functions.
\end{abstract}


\maketitle




%
%




\section{Introduction}\label{secVit.1}

In this paper I discuss some specific features of the formation of
topological defects during the process of phase transitions and the
relation between fractals and coherent states. Fractals and defects
are present in an extremely large number of systems and natural
phenomena and therefore much attention is devoted to their study.

Examples of defects are magnetic domain walls in ferromagnets,
vortices in superconductors and superfluids, dislocations, point
defects, etc. in crystals. Even in cosmology cosmic strings
\cite{kib} may be studied as topological singularities whose role
may have been relevant in the phase transition processes in the
early Universe. The analogy between defect formation in condensed
matter physics and high energy physics and cosmology is quite
surprising \cite{volovik1,volovik2,zurek,Bunkov}. Moreover, the
study of defect formation may be particularly useful in the
understanding of the same mathematical structure of quantum field
theory (QFT) with particular reference to the processes of
non-equilibrium symmetry breaking, a subject on which my attention
in this paper will be mostly focused.

On the other hand, there is no need of many words to express the
relevance of fractals in science, from physics to biology, medical
sciences, earth science, clustering of galaxies, etc. \cite{Bunde}.
It is therefore of great interest to investigate in deep fractal
properties. Here I present the functional realization of fractals in
terms of coherent states. From one side, this may shed some light on
the dynamical formation of fractals. From the other side, it also
exhibits the fractal nature of coherent states, thus opening new
perspectives in the analysis of those phenomena where coherent
states play a relevant role.

In my discussion I will closely follow some previous papers of mine
on similar topics, especially Refs. \cite{Links}-\cite{dice2005}. I
will essentially consider the symmetry properties of the dynamics of
the systems under consideration and therefore the conclusions will
be mostly model independent. On the contrary, a specific model
choice would imply working in some approximation scheme. The
general, more formal, approach I will follow is, however, more
convenient in discussing some of the general questions which arise
in the study of defect formation in the framework of non-equilibrium
symmetry breaking phase transitions.


Among the results that will be presented there is the theoretical
explanation, based on the microscopic dynamics of quantum fields, of
the fact that the formation of topological defects is typically
observed during the processes of non-equilibrium symmetry breaking
phase transitions when a gauge field is present and an order
parameter exists. The fact that the conclusions are, as said, model
independent is helpful since those features of defect formation are
singled out which are shared by many systems in a wide range of
energy scale, independently of specific aspects of the system
dynamics.

I will also present the proof that the Nambu-Goldstone (NG) particle
acquires an effective non-zero mass due to boundary (finite volume)
effects. This is related with the size of the defect and introduces
us to the question of the dynamical formation of the boundaries.
These can be considered indeed to be themselves like a "defect" of
the system, which otherwise, in the absence of boundaries, would
extend to infinity. An interesting question is indeed the one of the
formation of the system boundaries, which has to be, of course, a
{\it dynamical} process, not imposed by hand from the exterior
(presumably, such a dynamical formation of boundaries has a lot to
do with morphogenesis). As we will see, finite volume effects are
related with temperature effects.

Also the discussion of fractals will be of general character, aiming
to point out some structural fractal aspect rather than analyzing
the features of specific fractals. Indeed, I will focus my attention
only on a feature common to all the self-similar fractal structures.
The self-similar property of fractals is only one of the many
mathematical and phenomenological properties of fractals. It is,
however, a characterizing properties of an extremely large class of
fractals and therefore I focus my attention on it. The connection
will be made with the theory of the entire analytic functions and
with the $q$-deformed algebra of the (Glauber) coherent states. This
results in the possibility of incorporating fractal properties in
the framework of the theory of entire analytical functions.
Conversely, it also allows to recognize, in the specific sense that
will be discussed below (cf. Section \ref{secVit.5}), fractal
properties of coherent states. The study presented in this paper
thus provides a first step in the understanding of the  dynamical
origin of fractals and of their global nature emerging from local
deformation processes. It also provides insights on the geometrical
(fractal) properties of coherent states.

The presentation is organized as follows. In Section \ref{secVit.2}
I present the notion of spontaneous breakdown of symmetry (SBS) in
QFT.  As a result of SBS, the invariance properties of the basic
field equations manifest themselves in the formation of ordered
patterns at the level of the observable fields. We have thus the
phenomenon of the {\it dynamical rearrangement of symmetry}. In
Section \ref{secVit.3} I consider the problem of the formation of
'extended objects' (defects) with topological singularities in the
processes of phase transitions. It will be then evident the relevant
role played by the mathematical structure of QFT  characterized by
the existence of infinitely many unitarily inequivalent
representations of the canonical (anti-)commutation relations. The
effects of the non-vanishing mass of the Nambu-Goldstone (NG)
particles on the coherent domain size will be discussed in Section
\ref{secVit.4}, where temperature effects will be also considered.
The functional realization of fractals in the framework of the
theory of the entire analytical functions and their relation with
the deformed Weyl-Heisenberg algebra of the coherent states is
presented in Section \ref{secVit.5}. Section \ref{secVit.6} is
devoted to final remarks.

\section{Spontaneous breakdown of symmetry and
dynamical rearrangement of symmetry}\label{secVit.2}

In an intuitive picture, the formation of a topological defect may
be thought to occur when a spatially extended region of the "normal"
(i.e. symmetric) state is surrounded by "ordered" domains. In a
vortex, for example, the normal state constitutes the ``core'' of
the vortex which is ``trapped'' in an ordered state surrounding it.
Therefore, the first notions we need in order to study  defect
formation are the ones of ordered or non-symmetric state vs the
disordered or symmetric state. It is known that the mechanism of
spontaneous breakdown of symmetry in QFT
\cite{Umezawa:1993yq,Itzykson:1980rh} is at the origin of the
dynamical formation of ordered patterns out of the symmetric state.

In QFT \cite{Umezawa:1993yq,Itzykson:1980rh} the dynamics, i.e. the
Lagrangian and the nonlinear field equations from it derived, is
given in terms of the interacting fields, say $\varphi_{H} (x)$,
called the Heisenberg fields. The observables measured in
experiments are instead described in terms of asymptotic in- (or
out-) operator fields (called free or physical fields), say
$\varphi_{in} (x)$, which satisfy free field equations. Such a dual
level structure (Heisenberg or interacting fields vs free or
physical or asymptotic fields) is called the
Lehmann-Symanzik-Zimmermann (LSZ) standard formalism of QFT
\cite{Itzykson:1980rh}. In condensed matter physics and in
quark-gluon physics asymptotic in- (or out-) operator fields are not
available. In such cases the role of physical fields is played by
quasiparticle fields and by fields in the asymptotic freedom regime,
respectively. When it happens that the physical vacuum state of the
system is not symmetric under the action of one or more generators
of the group of transformations, say $G$, under which the Heisenberg
field equations are invariant, we say that the symmetry is
spontaneously broken. Here we are assuming that $G$ is a continuous
group of transformations.

Summarizing, we have the dynamics given in terms of Heisenberg
fields and the Hilbert space of physical states where asymptotic
fields, satisfying free field equations, are realized. The theory is
``solved" when the dynamical map between Heisenberg fields and
asymptotic fields is found; in other words, when the coefficients of
the map are computed by solving the Heisenberg field equations. The
dynamical map is a weak equality: it holds between expectation
values of the the members of the mapping computed in the Hilbert
space of the physical states. It is useful to proceed by considering
a concrete standard example, the one of the U(1) gauge model where
vortices appear. However, our conclusions apply, in their structural
features, to more complex models as well, including the non-Abelian
case of SO(3) and SU(2) symmetry (as for the monopole and the
sphaleron case \cite{Manka:1988tb}, respectively).

Let  ${\cal L}[\phi_{H}(x), {\phi_{H}}^*(x), A_{H \mu}(x)]$ be  the
Lagrangian density for a complex scalar field $\phi_{H}(x)$
interacting with a gauge field $A_{H \mu}(x)$. For our tasks, there
is no need to specify the detailed structure of the Lagrangian. We
only require that ${\cal L}$ be invariant under global and local
U(1) gauge transformations (the Higgs-Kibble model \cite{hig,ki}):
\bea\lab{lp1} \phi_{H}(x) &\rar& e^{i \te} \phi_{H} (x) ~, \qquad
\qquad A_{H \mu}(x) \rar A_{H \mu}(x), 
\\
\lab{lp2} \phi_{H} (x) &\rar& e^{i e_0 \la(x)} \phi_{H} (x) ~,
\qquad A_{H \mu} (x) \rar A_{H \mu} (x) \, + \, \pa_\mu \la(x), \eea
respectively. Here $\la(x)\rar 0$ for $|x_0|\rar \infty$ and/or
$|{\bf x}|\rar \infty$. The Lorentz gauge $\pa^\mu A_{H \mu}
(x)\,=\,0$ is adopted. Moreover, $\phi_{H} (x)\equiv
\frac{1}{\sqrt{2}}\left[\psi_{H} (x) + i \chi_{H} (x)\right]$.

We assume that spontaneous breakdown of global U(1) symmetry can
occur, i.e. that $\lan 0| \phi_H(x)|0\ran \equiv {\ti v} \neq 0$.
${\ti v}$ is a constant and $\rho_{H} (x) \equiv \psi_{H} (x) - {\ti
v}$.

A crucial point is that in the presence of SBS the theory contains a
massless negative norm field (ghost) $b_{in} (x)$, the
Nambu-Goldstone massless mode $\chi_{in} (x)$, and a massive vector
field $U^\mu_{in}$, as one can show, e.g. by functional integration
techniques \cite{MPUV75}.

We can then exhibit a concrete example of the LSZ mapping (the
dynamical map, also called the Haag expansion) between interacting
fields and physical fields, mentioned above. It provides the
relation between the dynamics and the observable properties of the
physical states. The following LSZ maps are indeed found
\cite{MPUV75}:
\bea \non 
\phi_H(x)&=& :\exp\left\{i
\frac{Z_\chi^{\frac{1}{2}}} {{\ti v}}\chi_{in}(x) \right\}
\left[{\ti v} + Z_\rho^{\frac{1}{2}} \rho_{in}(x) + F[\rho_{in},
U^\mu_{in}, \pa(\chi_{in} - b_{in})] \right]: 
\\
\lab{lp57b} A^{\mu}_{H}(x) &=& {Z_3^{\frac{1}{2}}} U^{\mu}_{in}(x)+
{\frac{Z_\chi^{\frac{1}{2}}}{e_0{\ti v}}} \pa^\mu b_{in}(x)+ :
F^{\mu}[\rho_{in}, U^\mu_{in}, \pa(\chi_{in}-b_{in})]:. \eea
The colon symbol denotes normal ordering. $ Z_\chi $, $ Z_\rho $ and
$Z_3$ are the wave function renormalization constants. The
functionals $F$
and $F^{\mu}$
are determined when a specific explicit choice for the Lagrangian is
assumed. In the present case, the above results are model
independent. When considering a specific choice for the Lagrangian
they can be also obtained within some approximation, e.g., by using
the saddle point expansion.

Another important relation is the one providing the $S$-matrix. This
is given by
\be \label{SM} S\,=\,: S[\rho_{in}, U^\mu_{in}, \pa(\chi_{in} -
b_{in})]: ~.\ee
I stress that, as already observed above, the dynamical mappings
(\ref{lp57b}) and (\ref{SM})  are weak equalities, i.e. they are
equalities among matrix elements computed in the Fock space for the
physical states. The free field equations are
\be\lab{lp24} \pa^2 \chi_{in}(x)\,=\,0~, \quad \pa^2
b_{in}(x)\,=\,0~, \quad (\pa^2 \, + \, m_\rho^2)\rho_{in}(x) \,
=\,0~, \ee
\be\lab{lp29} ( \pa^2 \, + \, {m_V}^{2}) U^\mu_{in}(x) \, =\, 0 ~,
\qquad \pa_{\mu} U^\mu_{in}(x) \, =\, 0~. \ee
with ${m_V}^{2} = \frac{Z_{3}}{Z_\chi} (e_0{\ti v})^{2}$.

I now observe that the global gauge transformations (\ref{lp1}) of
the Heisenberg fields are found to be induced by the in-field
transformations (see (\ref{lp57b}) and (\ref{SM})):
\be\lab{lp53a} \chi_{in}(x)  \rar  \chi_{in}(x) \, + \, \frac{{\ti
v}}{Z_\chi^{\frac{1}{2}}} \te f(x)   ~, \ee
\be\lab{lp53b} b_{in}(x) \rar b_{in}(x), \quad \rho_{in}(x) \rar
\rho_{in}(x), \quad U^\mu_{in}(x) \rar U^\mu_{in}(x) ~. \ee
The local phase transformations (\ref{lp2}) are induced by
\be\lab{lp51a} \chi_{in}(x)  \rar   \chi_{in}(x) \, + \, \frac{e_0
{\ti v}}{Z_\chi^{\frac{1}{2}}} \la(x) ~, \qquad b_{in}(x)  \rar
b_{in}(x) \, + \, \frac{e_0 {\ti v}}{Z_\chi^{\frac{1}{2}}} \la(x)~,
\ee
\be\lab{lp51c} \rho_{in}(x)  \rar   \rho_{in}(x)~, \qquad
U^\mu_{in}(x) \rar  U^\mu_{in}(x)  ~. \ee

An important remark is that (\ref{lp53a}) with $f(x)= 1$
(translation by a constant c-number) is not unitarily implementable
in QFT (i.e. in the limit of infinitely many degrees of freedom):
$f(x)$ is introduced in order to make the generator of such a
transformation well defined. Mathematical definiteness requires that
$f(x)$  be a square integrable function, solution of the equation
for $\chi_{in}(x)$ and $b_{in}(x)$, i.e. $\pa^2 f(x) =0$. The limit
$f(x)\rar 1$ (i.e. the infinite volume limit) is to be performed at
the end of the computation. The in-field equations and the $S$
matrix are invariant under the above in-field transformations (in
the limit $f \rar 1$).

Notice that the generator (the Glauber's displacement operator) of
the transformation (\ref{lp53a}) where $f(x)$ is set to be $1$,
namely the translation (or displacement or shift) of the operator
field $\chi_{in}(x)$ by a constant, is the key ingredient in the
theory of coherent states in quantum mechanics (QM). There, since
the number of the degrees of freedom is finite (finite volume) the
von Neumann theorem guaranties that the representations of the
canonical commutation relations are unitarily equivalent. In
contrast, in QFT  infinitely many unitarily inequivalent
representations exist \cite{VonNeumann:1931a}. The {\it possibility}
of choosing one of them among the many inequivalent ones corresponds
to a completely non-trivial condition under which the basic dynamics
is realized in terms of observable quantities.


Let me now briefly comment on the physical meaning of the formal
structure which has emerged as a result of SBS.

We have seen that the global and the local U(1) gauge
transformations of the Heisenberg fields are induced, at the level
of the physical fields, by the group $G'$ of transformations
(\ref{lp53b}) and (\ref{lp51a}), respectively. Since $G'$ is a
different group of transformation than $G$, one says that $G \rar
G'$ represents {\it the dynamical rearrangement of symmetry}
\cite{Umezawa:1993yq,Um1}: it is the result of, and expresses the
consistency between the invariance of the Lagrangian under the $G$
symmetry transformations and the SBS condition (in the considered
model, $G=$ U(1) and the SBS condition is $\lan 0| \phi_H(x)|0\ran =
{\ti v} \neq 0$).

Translations of boson fields (\ref{lp53a}) and  (\ref{lp51a}) are
thus obtained as a consequence of SBS. $G'$ is the {\it group
contraction} of the U(1) symmetry group of the dynamics; if the
dynamics symmetry group is SU(n) or SO(n), the group contraction
$G'$ is EU(n-1) or E(n), respectively \cite{inonu}. Under quite
general conditions, the dynamical rearrangement of symmetry is found
to lead to the group contraction of the group under which the
Lagrangian is invariant. This is an exact result which goes beyond
any approximation scheme \cite{inonu}. I stress that $G'$ is the
transformation group relevant to the phase transitions process.

\subsection{Homogeneous and non/homogeneous boson
condensation}\label{secVit.21}

``Shifts'' of the NG fields are thus introduced through the
dynamical rearrangement of symmetry. They are controlled by the
Abelian subgroup of $G'$. In the global gauge case, the
transformation (\ref{lp53a}), with $f(x)= 1$,  describes the NG {\it
homogeneous boson condensation}. As a result of such a boson
condensation in the ground state, coherent long range correlation is
established which manifests as the ordered pattern in the system
ground state. We thus realize that transitions between the system
phases characterized by different ordered patterns in the ground
state may be induced by the process of boson condensation. Stated in
different words: phase transitions are induced by variations
(gradients) of the NG boson condensation. Since in the $f(x)= 1$
limit it does not exist any unitary generator of the boson
translation by a constant c-number, vacua with different boson
condensation are unitarily inequivalent states and thus we see that
phase transitions are transitions among unitarily inequivalent Fock
spaces. The essentially non-perturbative nature of the phase
transition process is in this way recognized. Moreover, we learn
that since phase transitions can only occur when a multiplicity of
unitarily inequivalent spaces of states is available, they can only
occur in QFT. In quantum mechanics, indeed, all the representations
of the canonical commutation relations are unitarily equivalent, as
stated by the von Neumann theorem, and therefore no (phase)
transition between inequivalent state spaces is conceivable.

When $f(x) \neq 1$ the NG translation (\ref{lp53a}) describes
coherent {\it non-homogeneous boson condensation} and (\ref{lp53a})
is called the {\it boson transformation}. As we will see below,
extended objects (defects) are described by non-homogeneous boson
condensation.

Next step is to consider ``macroscopic'' manifestations of the boson
condensation. In order to do that, I observe that in the framework
of the U(1) gauge model discussed above, the Maxwell equations are
given by
\be\lab{lp37} - \pa^2 A_{H}^{\mu}(x) \, =\, j_{H}^{\mu}(x) \, -\,
\pa^{\mu} B(x) ~, \ee
where
\be\lab{lp43} B(x)= \frac{e_0 {\ti
v}}{Z_\chi^{\frac{1}{2}}}[b_{in}(x) - \chi_{in}(x)]  \, ~, \quad
\pa^2 B(x)\, =\,0   ~, \ee
and $j_{H}^{\mu}(x)= \de{\cal L}(x)/\de A_{H \mu} (x)$. Let the
current $j_{H}^{\mu}$ be the only source of $A_{H}^{\mu}$ in any
observable process. This implies  $_p\lan b|\pa^{\mu} B(x)|a\ran_p\,
= \,0$, i.e.
\be\lab{lp45} (- \pa^2) \,_p\lan b| {A^{0 \mu}}_{H}(x) |a \ran_p \,
= \,_p\lan b| j_{H}^{\mu}(x) |a\ran_p  ~, \ee
where $A^{0\mu}_{H}(x) \equiv A^{\mu}_{H}(x) - { e_0{\ti v}}:\pa^\mu
b_{in}(x):$. $|a\ran_p $ and $|b\ran_p $ are two generic physical
states. The condition $_p\lan b|\pa^{\mu} B(x)|a\ran_p\, = \,0$
leads to the Gupta-Bleuler-like condition
\be\lab{lp49} [\chi_{in}^{(-)}(x)  \, - \,
b_{in}^{(-)}(x)]|a\ran_p\, = \,0  ~, \ee
where $\chi_{in}^{(-)}$ and $b_{in}^{(-)}$ are the
positive-frequency parts of the corresponding fields. Eq.
(\ref{lp49}) shows that $\chi_{in}$ and $b_{in}$ do not participate
to any observable reaction. However,  the NG bosons do not disappear
from the theory: their condensation in the vacuum can have
observable effects.

Remarkably, {\it Eq.(\ref{lp45}) is the classical Maxwell
equations}.

The boson transformation must be also compatible with the physical
state condition (\ref{lp49}). $B$ changes as
\be\lab{vs9} B(x) \rar B(x) - \frac{e_0 {\ti v}^2}{Z_\chi} f(x) ~
\ee
under the transformation $\chi_{in}(x)  \rar  \chi_{in}(x) +
\frac{{\ti v}}{Z_\chi^{\frac{1}{2}}} f(x)$. Eq. (\ref{lp45}) is then
violated. In order to restore it, the shift in $B$ must be
compensated by means of the transformation on $U_{in}$:
\be\lab{vs10} U^{\mu}_{in}(x) \rar U^{\mu}_{in}(x) +
{Z_{3}}^{-\frac{1}{2}} a^{\mu}(x) \qquad , \qquad \pa_\mu
a^{\mu}(x)=0 ~, \ee
with a convenient c-number function $a^{\mu}(x)$. The dynamical maps
of the various Heisenberg operators are not affected by (\ref{vs10})
provided
\be\lab{Bvs20} (\pa^2 + m_V^2) a_\mu(x) \, = \,\frac{m_V^2}{ e_0}
\pa_\mu f(x)~. \ee
{\it Eq. (\ref{Bvs20}) is the classical Maxwell equation for the
vector potential $a_{\mu}$} \cite{MPUV75,MPU75}. Thus we see that
symmetry breaking phase transitions are characterized by macroscopic
ground state effects, such as the vacuum current and field (e.g. in
superconductors), originated from the microscopic dynamics.

These results are not confined to the U(1) model here considered.
They are also obtained in more complex models with non-Abelian
symmetry groups (see \cite{Manka:1988tb,Um1}), in the relativistic
as well as in non-relativistic regime.

\section{Defect formation and phase transitions}\label{secVit.3}

We have seen that non-homogeneous boson condensation is a strict
consequence of SBS provided certain conditions are met ($f(x) \neq
1$). Next question is why defect formation is observed during the
processes of non-equilibrium symmetry breaking phase transitions
when a gauge field is present and an order parameter exists. To
answer to such a question from the perspective of the microscopic
dynamics, we need to consider the topological characterization of
non-homogeneous boson condensation in SBS theories \cite{les}.

The boson transformation function $f(x)$ considered in the previous
Section plays the role of a "form factor": the extended object (the
defect) appears as the macroscopic envelope of the non-homogeneous
boson condensate localized over a finite domain. The topological
charge of the defect is thus expected to arise from the topological
singularity of the boson condensation function.

The boson condensation has been recognized in Section \ref{secVit.2}
to be formally obtained by translations of boson fields, say
$\chi_{in}(x) \rar \chi_{in} (x) + f(x)$, with c-number function
$f(x)$, satisfying the same field equation for $\chi_{in} (x)$. As
already said, these translations are called {\it boson
transformations} \cite{Umezawa:1993yq}. According to its general
definition, the boson transformation may be applied to boson fields
which need not to be necessarily massless.

We have also seen that transitions between phases characterized by
different ordered patterns in the ground state are induced by
variations (gradients) of the boson transformation function. Thus,
in order to show that in such a transition process the conditions
can be met for the formation of topological defects, we need to
consider under which constraints the boson transformation function
$f(x)$ can carry a topological singularity. Then we have also to
show that these constraints are in fact satisfied in the process of
phase transitions.

With respect to the first of these points, one can show that
topological singularities of the boson transformation functions are
allowed only for for massless bosons \cite{Umezawa:1993yq,Um1}, such
as NG bosons of SBS theories where ordered ground states appear.

In the boson transformation $\chi_{in}(x)  \rar  \chi_{in}(x) +
f(x)$, let $f(x)$ carry a topological singularity. This means that
it is path-dependent:
\be
\\ \lab{ts1}
G^{\dag}_{\mu\nu}(x) \equiv [\pa_\mu,\pa_\nu]\,f(x) \neq 0~, \qquad
{\rm for \;certain} \quad \mu\, , \,  \nu \, , \, x  ~. \ee
$\pa_\mu \, f$ is related with observables  (see below) and
therefore it assumed to be single-valued, i.e.
$[\pa_\rho,\pa_\nu]\,\pa_\mu f(x)\,=\,0$. $f(x)$ is required to be
solution of the $\chi_{in}$ equation. Suppose that in such an
equation there is a non-zero mass term: $(\pa^2 + m^2)f(x) = 0$.
From the regularity of $\pa_\mu f(x)$ it follows that
\be\lab{ts5} \pa_\mu f(x) \, =\, \frac{1}{\pa^2 \, + \, m^2}
\pa^{\la} \,G^{\dag}_{\la\mu}(x) ~, \ee
which leads to $\pa^2 f(x) \, =\,0$, which in turn implies $m=0$.
Thus (\ref{ts1}) is  compatible only with massless $\chi_{in}$. This
explains why topological defects are observed only in systems
exhibiting massless modes, such as ordered patterns, namely in the
presence of NG bosons sustaining long range ordering correlation.

For the second point, I recall that Eq. (\ref{Bvs20}) is a
characterizing equation for the occurrence of the phase transition
processes. From such an equation we see that the classical ground
state current $j_{\mu}$ is given by
\be\lab{vs21} j_\mu(x)\equiv \lan 0| j_{H \mu}(x) |0 \ran \, =\,
 m_V^2 \left[ a_\mu(x) - \frac{1}{e_0} \pa_\mu f(x) \right]~.
\ee
Here $ m_V^2  a_\mu(x)$ is the {\em Meissner-like current} and $
\frac{m_V^2}{e_0} \pa_\mu f(x)$ is the {\em boson current}.

Eq. (\ref{vs21}) shows that the classical field and the classical
current do not occur for regular $f(x)$ ($G^{\dag}_{\mu\nu} = 0$),
with $\pa^{2}f(x) = 0$ (which is required when considering NG
bosons).

In fact, from (\ref{Bvs20}) $a_\mu$ is formally given by $a_\mu(x)=
 \frac{1}{\pa^{2} + m_V^2 } \frac{m_V^2}{e_0} \pa_\mu f(x)$. Then we have
 ${\pa^2} a_\mu(x)= \pa^{2} \frac{1}{\pa^{2} + m_V^2 }
\frac{m_V^2}{e_0} \pa_\mu f(x)= 0$ for regular $f(x)$, i.e.
$a_{\mu}(x) = \frac{1}{e_{0}} \pa_{\mu} f(x)$. Thus we see that for
regular $f(x)$ the Meissner-like current and the boson current
cancel each other, which implies zero classical current ($j_{\mu} =
0$) and zero classical field ($F_{\mu\nu} = \pa_{\mu} a_{\nu} -
\pa_{\nu} a_{\mu}$). The gauge potential behaves thus as the
``reservoir'' compensating the boson transformation gradients
\cite{nakamura}.

It is interesting to observe that the gauge field potential
$a_{\mu}(x)$ can be thought as ``generated'' by the gradient
$\frac{1}{e_{0}} \pa_{\mu} f(x)$ of the boson condensation of the
$\chi_{in}(x)$ field. Its introduction in Eq. (\ref{vs10}) was
indeed motivated by the stability requirement of the physical state
$|a \rangle_p$ under the shift of the $\chi_{in} (x)$ field. For
regular $f(x)$, $a_{\mu}(x)$ exactly compensates $\frac{1}{e_{0}}
\pa_{\mu} f(x)$. For singular $f(x)$, the non-vanishing difference
${m_{V}^2}[a_{\mu}(x) - \frac{1}{e_{0}} \pa_{\mu} f(x)] \equiv
j_\mu(x)$  satisfying the continuity equation $\pa^{\mu}j_\mu(x) =
0$ behaves as a current (source) in the Maxwell equation
(\ref{Bvs20}).

In conclusion, vacuum currents characterizing the processes of phase
transition appear only when $f(x)$ has topological singularities,
which, as we have seen above, is only compatible with the
condensation of massless bosons, as it happens  when SBS occurs.

Summarizing, the same conditions allowing topological singularities
in the boson condensation function $f(x)$ are the ones under which
phase transitions may occur in a gauge theory. Therefore, the
conditions for the formation of topological defects are met in the
phase transition processes, which explains why topological defects
are observed in the process of symmetry breaking phase transitions.

I note that the assumption of the regularity of ${\pa}_{\mu}f$ is
justified by the (topological) regularity of observable quantities.
The  classical current (\ref{vs21}), which is an observable
quantity, is indeed given in terms of gradients ${\pa}_{\mu}f$ of
the boson condensation function.

Notice that the appearance of space-time dependent order parameter
$\ti v$ is not enough to guarantee that persistent ground state
currents (and fields) will exist. Indeed, if $f$ is a regular
function, the space-time dependence of $\ti v$ can be gauged away by
an appropriate gauge transformation. Therefore, topological defects
cannot be obtained in such case. I also note that in a theory which
has only global gauge invariance non-trivial physical effects, like
linear flow in superfluidity, may be produced by non-singular boson
transformations of the NG fields.

We may also discuss the effects of topological singularity in the
$S$ matrix. Since  boson transformations with regular $f$ do not
affect observable quantities, the $S$ matrix must be actually given
by
\be\lab{lp56aa} S\,=\,: S[\rho_{in}, U^\mu_{in} - \frac{1}{m_V}
\pa(\chi_{in} -
 b_{in})] : ~,
\ee
which is in fact independent of the boson transformation with
regular $f$:
\be\lab{lp56ab} S\,\rar\,S' = : S[\rho_{in}, U^\mu_{in} -
\frac{1}{m_V} \pa(\chi_{in} - b_{in}) + Z^{-\frac{1}{2}}_{3}
(a^{\mu} - \frac{1}{e_{0}} {\pa}^{\mu} f)]: \ee
since  $a_{\mu}(x) = \frac{1}{e_{0}} \pa_{\mu} f(x)$ for regular
$f$. However, $S' \neq S$ for singular $f$: in such a case
(\ref{lp56ab}) shows that $S'$ includes the interaction of the
quanta $U^\mu_{in}$ and $\rho_{in}$ with the classical field and
current. This shows how it may happens that {\it quanta interact and
have effects on classically behaving macroscopic extended objects}.

The above conclusions are not limited by dimensional considerations
or by the Abelian or non-Abelian nature of the symmetry group. They
apply to a full set of topologically non-trivial extended objects,
such as topological line singularity,  surface singularity, grain
boundaries and dislocation defects in crystals, SU(2)-triplet model
and monopole singularity. The topological singularity and the
topological charge of the related extended object can be completely
characterized. A detailed account can be found in Refs.
\cite{Um1,les}. The general character of our conclusions also shows
why the features of the defect formation are shared by quite
different systems, from condensed matter to cosmology. They account
for the macroscopic behavior of extended objects and their
interaction with quanta in a unified theoretical scheme. In the
following Section I consider finite volume and temperature effects
in such a scheme.

\section{Defect formation and non-vanishing effective mass
of the Nambu-Goldstone bosons}\label{secVit.4}

The two-point function of the $\chi(x)$ field in the considered U(1)
model can be computed in full generality by using the Ward-Takahashi
identities. It has \cite{MPUV75} the following pole structure for :
\be\lab{lp160}
 \lan \chi(x)\chi(y)\ran = \lim_{\ep\rar 0} \left\{
\frac{i}{(2\pi)^4} \int d^4p\, \frac{Z_{\chi} e^{-i p (x-y)}}{p^2 -
{m_{\chi}}^{2}+i \ep a_\chi} + cont.~c. \right] , \ee
where $Z_\chi$ and $a_\chi$ are renormalization constants and
'cont.~c.' denote continuum contributions. The space integration of
$\lan \chi(x)\chi(y)\ran$ picks up the pole contribution at $p^2=0$,
and leads to \cite{MPU741,MPU742}
\be\lab{gp9} {\ti v}= \frac{Z_\chi}{a_\chi} v \Lrar m_\chi = 0 ~,~~~
or ~ ~~ {\ti v}= 0 \Lrar m_\chi \neq 0   ~, \ee
where $v$ denotes a convenient c-number \cite{MPUV75}. Eq.
(\ref{gp9}) proves the existence of a massless particle
corresponding to the pole singularity. It expresses the well known
Goldstone theorem: if the symmetry is spontaneously broken (${\ti v}
\neq 0$), the NG massless mode exists, whose interpolating
Heisenberg field is $\chi_{H} (x)$. It spans the whole system since
it is massless and manifests as a long range correlation mode. Thus
it is responsible for the vacuum ordering.

I restrict now the space integration of Eq. (\ref{lp160}) over the
finite (but large) volume $V \equiv \eta ^{-3}$. For each space
component of $p$ we have:
\be\lab{note13}
\de_\eta(p)=\frac{1}{2\pi}\int_{-\frac{1}{\eta}}^{\frac{1}{\eta}}
dx\,e^{ipx}=\frac{1}{\pi p}\, sin\frac{p}{\eta}  ~. \ee
As well known,  $ lim_{\eta\rar 0} \de_\eta(p)=\de (p) $ and
\be\lab{note15} lim_{\eta\rar 0} \int dp\, \de_\eta(p)\,f(p)=f(0)=
lim_{\eta\rar 0} \int dp\, \de(p-\eta)\,f(p)  ~. \ee
Using $\de_\eta(p)\simeq\de(p-\eta)$ for small $\eta$, one obtains
\be\lab{6} {\ti v}(y,\ep , \eta)=i{\ep} v
e^{-i\vec{\eta}\cdot\vec{y}}\, \Delta_{\chi}(\ep,\vec{\eta},p_0=0)
~, \ee
where
\be\lab{gp8} \De_\chi(\ep, \vec{\eta}, p_{0}=0) = \left[
\frac{Z_\chi}{{- \omega_{\vec{p} = \vec{\eta}}^{2}} +i\ep  a_\chi} +
(continuum \;contributions) \right]\, , \ee
and ${\omega^{2}}_{\vec{p} = \vec{\eta}} = \vec{\eta}^{2} +
{m_{\chi}}^{2}$. Thus, ${lim_{\ep \rar 0}} {lim_{\eta \rar 0}} {\ti
v}(y, \ep, \eta) \neq 0$ only if $m_{\chi} = 0$, otherwise ${\ti v}
= 0$. The Goldstone theorem is of course recovered in the infinite
volume limit ($\eta \rar 0$)  (the QFT limit).

Note that if $m_{\chi} = 0$ and $\vec{\eta}$ is given a non-zero
value (i.e. by reducing to a finite volume, i.e. in the presence of
boundaries), then ${\omega}_{\vec{p} = \vec{\eta}} \neq 0$ and it
acts as an "effective mass" for the $\chi$ bosons. Then, in order to
have the order parameter ${\ti v}$ different from zero $\ep$ must be
kept non-zero. I remark that an impurity embedded in the system
always generates ``boundaries'' around it, thus producing finite
volume effects.

In conclusion, near the boundaries ($\eta \neq 0$) the NG bosons
acquire an effective mass $m_{eff} \equiv {\omega}_{\vec{p} =
\vec{\eta}}$. Then they propagate over a range of the order of
$\xi\equiv\frac{1}{\eta}$, which is the linear size of the
condensation domain, or, in the presence of topological singularity,
the size of the topologically non-trivial condensation, namely of
the extended object (the defect). It must be observed that the
topological singularity tends to be washed out since, according to
the conclusion of the previous Section, it is not compatible with
the non-vanishing value of the NG boson effective mass. Near the
boundaries we thus expect (topologically) regular boson
condensation. Far from the boundaries, topological singularities
might survive.

If  $\eta \neq 0$ then $\ep$ must be non-zero in order to have the
order parameter different from zero, ${\ti v} \neq 0$ (at least
locally). In such a case the symmetry breakdown is maintained
because $\ep \neq 0$: $\ep$ acts as the coupling with an external
field (the pump) providing energy. Energy supply is required in
order to condensate modes of non-zero lowest energy
${\omega}_{\vec{p} = \vec{\eta}}$. Boundary effects are thus in
competition  with  the breakdown of symmetry \cite{les}. They may
preclude its occurrence or, if symmetry is already broken, they may
reduce to zero the order parameter.

The above discussion fits with the intuitive picture: for large but
finite volume one expects that the  order parameter is constant
``inside the bulk'' {\em far} from the boundaries. However, ``{\em
near}'' the boundaries, one might expect ``distortions'' in the
order parameter: ``near'' the system boundaries we may have
non-homogeneous order parameter, ${\ti v} ={\ti v}(x)$ (or even
${\ti v}\rar 0$). Such non-homogeneities in the boson condensation
``smooths out'' in the $V\rar\infty$ limit.

Remarkably, the r\^oles of the boundaries and of the NG effective
mass in the above discussion may be exchanged, i.e. if for some
kinematical or dynamical reasons the NG modes acquire non-vanishing
effective mass, say of the order of $\eta$, then the ordered domains
will have linear dimension of the order of $\frac{1}{\eta}$, which
means that the domain  boundaries  are {\it dynamically generated}.

I observe that by use of a model system where two level atoms are
considered in interaction with their radiative field, the analysis
of stability of the solutions of field equations shows
\cite{DelGiudice:2006a} that the e.m. field, as an effect of the
spontaneous breakdown of the phase symmetry, gets a massive
component (the amplitude field), as indeed expected in the
Anderson-Higgs-Kibble mechanism \cite{MPUV75}, there is a
(surviving) massless component (the phase field) playing the role of
the NG mode and the stability regime is reached provided the {\it
phase locking} of the e.m. and matter fields is attained. The
physical meaning of the phase locking can be stated as follows. The
gauge arbitrariness of the field $a_{\mu} (x)$ is meant to
compensate exactly the arbitrariness of the phase of the matter
field in the covariant derivative $D_{\mu} = \pa_{\mu} - i
ga_{\mu}(x)$. Should one of the two arbitrariness be removed by the
dynamics, the invariance of the theory requires the other
arbitrariness, too,  must be simultaneously removed, namely the
appearance of a well defined phase of the matter field implies that
a specific gauge function must be selected. The above link between
the phase of the matter field and the gauge of $a_{\mu}(x)$ is
stated by the equation $a_{\mu}(x) \propto \pa_{\mu} f(x)$
($a_{\mu}(x)$ is a pure gauge field) and the analysis above reported
in connection with the (topological) regularity and singularity of
$f(x)$ is then recovered.

\subsection{Remark on temperature effects and critical
regime}\label{secVit.41}

Let me briefly comment now on the temperature  effects on the order
parameter (symmetry may be restored at or above a critical
temperature $T_{C}$). See \cite{Manka:1988tb,Manka:1986a} for
further details.

Since the order parameter goes to zero when NG modes acquire
non-zero effective mass (unless, as observed above, external energy
is supplied), the effect of thermalization may be represented in
terms of finite volume effects by putting, e.g., $\eta \propto
{\sqrt{\frac{|T-T_C|}{T_C}}}$, so that temperature fluctuations
around $T_{C}$ may produce fluctuations in the size $\xi$ of the
condensed domain.

At $T > T_{C}$, but near to $T_{C}$, and in the presence of an
external driving field ($\ep \neq 0$), one may have the formation of
ordered domains of size $\xi \propto
({\sqrt{\frac{|T-T_{C}|}{T_C}}})^{- 1}$ even {\it before} transition
to fully ordered phase is achieved as $T \rar T_{C}$. As far as
$\eta \neq 0$, the ordered domains (and the topological defects) are
unstable. They disappear as the external field coupling $\ep \rar
0$. If ordered domains are still present at $T<T_C$, they also
disappear as $\ep\rar 0$. The surviving possibility of such ordered
domains below $T_C$ depends on the speed at which $T$ is lowered
(which is related to the speed at which $\eta \rar 0$), compared to
the speed at which the system is able to get homogeneously ordered.
The system is said to be in the critical or Ginzburg regime during
the lapse of time in which a maximally stable new configuration is
attained since the transition has started. In many cases,
information on the critical regime behavior is provided by using the
harmonic approximation for the evolution of the order parameter
$\tilde v (x)$ (non-homogeneous condensate)
\cite{zurek,Alfinito:2001aa,Alfinito:2001mm}. The reality condition
on the 'mass parameter' $M_{k}(t)$  turns out to be a condition on
the $k$-modes propagation. The ``effective causal horizon"
\cite{kib2,zurek1} can happen to be inside the system (possible
formation of more than a domain) or outside (single domain
formation) according to whether the time occurring for reaching the
boundaries of the system is longer or shorter than the allowed
propagation time. This determines the dimensions to which the
domains can expand. The number of defects (of vortices) $n_{def}$
possibly appearing during the critical regime and the evolution of
the size of the domain can be computed
\cite{Alfinito:2001mm,kib2,zurek1}.

It can be shown that higher momentum modes survive longer, which
implies that smaller size domains are more stable than larger size
domains \cite{Alfinito:2001mm,zurek1}. Correlation modes with
non-vanishing effective mass thus generate domains which tend to
break down into smaller, more stable domains. It might be worth to
investigate deeper such an occurrence since it might shed some light
also on macroscopic phenomena in biology, finance, etc., in view of
the fact that coherent condensed states are nearest to classical
states (we have seen in Sections \ref{secVit.2} and \ref{secVit.3}
how classical motion equations and classical observables are
obtained from the microscopic analysis). As an example, the above
analysis might suggest, within proper conditions and under
convenient extrapolations, that a global market may be maintained
only provided an (enormous) amount of energies is pumped in; its
``natural''  destiny being, otherwise, its breakdown into separate
``pieces''.

\section{Fractals and the algebra of coherent states}\label{secVit.5}

In this Section I show that a relation between fractals and the
algebra of coherent states exists (I have conjectured the existence
of such a relation in Ref. \cite{dice2005}).

In the following I consider the case of fractals which are generated
iteratively according to a prescribed recipe, the so-called
deterministic fractals (fractals generated by means of a random
process, called ``random fractals'' \cite{Bunde}, will be considered
in a future work).

I will focus my discussion on  the self-similarity property which is
in some sense the {\it most important property} of fractals (p. 150
in Ref. \cite{Peitgen}).

To be specific, I consider the {\it Koch curve} (Fig. 1). One starts
with a one-dimensional, $d = 1$, segment $u_0$ of unit length $L_0$,
sometimes called the {\it initiator} \cite{Bunde}. I call this, as
usually done, the step, or stage, of order $n = 0$. The length $L_0$
is then divided by the reducing factor $s =3$, and the rescaled unit
length $L_1 = \frac{1}{3} L_0$ is adopted to construct the new
``deformed segment'' $u_1$ made of $\alpha = 4$ units $L_1$ (step of
order $n = 1$).  $u_1$ is called the {\it generator} \cite{Bunde}.
Of course, such a ``deformation'' of the $u_0$ segment is only
possible provided one ``gets out'' of the one dimensional straight
line $r$ to which the $u_0$ segment belongs: this suggests that in
order to construct the $u_1$ segment ``shape'' the one dimensional
constraint $d = 1$ is relaxed. We thus see that such a shape, made
of $\alpha = 4$ units $L_1$, lives in some $d \neq 1$ dimensions and
thus we write $u_{1,q}(\alpha) \equiv q \, \alpha \, u_0$, $~q  =
\frac{1}{3^d}$, $~d \neq 1$, where $d$ has to be determined and the
index $q$ has been introduced in the notation of the deformed
segment $u_{1}$.

\vspace{.6cm}

\centerline{\epsfysize=1.8truein\epsfbox{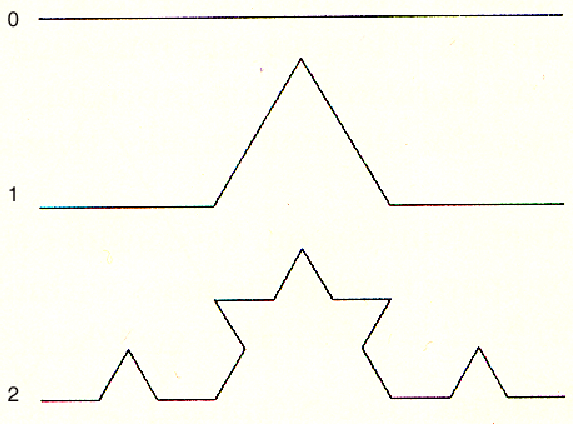}}
\vspace{.2cm}

\centerline{\epsfysize=1.8truein\epsfbox{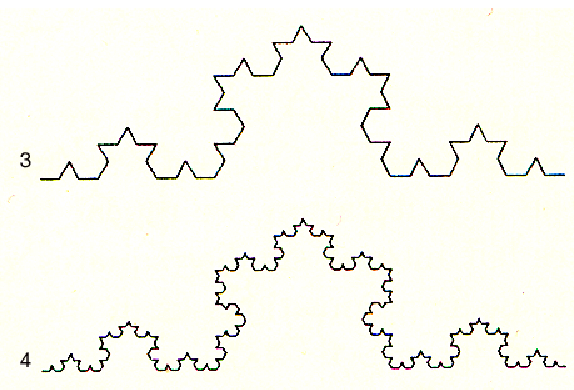}} \vspace{.2cm}

\centerline{{\small \noindent Fig. 1. The first five stages of Koch
curve.}}

\vspace{.6cm}

Usually one considers the familiar scaling laws of lengths, surfaces
and volumes when lengths are (homogeneously) scaled. Denoting by
${\cal H}(L_0)$ lengths, surfaces or volumes one has
\be \lab{B} {\cal H}(\la L_0) = \la^{d} {\cal H}(L_0) ~, \ee
under the scale transformation: $L_0 \rar \la L_0$.  A square $S$
whose side is $L_0$ scales to $\frac{1}{2^2}S$ when $L_0 \rar \la
L_0$  with $\la = \frac{1}{2}$. A cube $V$ of same side with same
rescaling of $L_0$ scales to $\frac{1}{2^3}V$. Thus $d = 2$ and $d =
3$ for surfaces and volumes, respectively. Note that
$\frac{S(\frac{1}{2}L_0)}{S(L_0)} = p =\frac{1}{4}$ and
$\frac{V(\frac{1}{2}L_0)}{V(L_0)} = p = \frac{1}{8}$, respectively,
so that in both cases $p = \la^{d}$. Similarly, for the length $L_0$
it is $p = \frac{1}{2} = \frac{1}{2^{d}} = \la^{d}$ and of course it
is $d = 1$.

By generalizing and extending this to the case of any other
``ipervolume'' ${\cal H}$ one considers thus the ratio
\be \lab{Ba} \frac{{\cal H}(\la L_0)}{{\cal H}(L_0)} = p ~, \ee
and assuming that Eq. (\ref{B}) is still valid ``by definition'',
one obtains
\be \lab{Bab} p ~{\cal H}(L_0) = \la^{d} {\cal H}(L_0)  ~, \ee
i.e. $p = \la^{d}$. Going back then to the discussion of the Koch
curve and setting $\al = \frac{1}{p} = 4$ and $q = \la^{d} =
\frac{1}{3^d} $, the relation $p = \la^{d}$ gives
%
%
\be \lab{30a} q \alpha = 1~, \quad {\rm where} \quad \al = 4, \quad
q= \frac{1}{3^d}~,\ee
i.e.
\be \lab{30} d = \frac{\ln 4}{\ln 3} \approx 1.2619~. \ee
The non-integer $d$ is called the {\it fractal dimension}, or the
{\it self-similarity dimension} \cite{Peitgen}.

With reference to the Koch curve, I observe that the meaning of Eq.
(\ref{Bab}) is that in the ``deformed space'', to which $u_{1,q}$
belongs, the set of four segments of which $u_{1,q}$ is made
``equals'' (is equivalent to) the three segments of which $u_0$ is
made in the original ``undeformed space''. The (fractal) dimension
$d$ is the dimension of the deformed space which allows the
possibility of such an ``equivalence'', i.e. that ensures the
existence of a solution of the relation $\frac{1}{\al} =\frac{1}{4}
= \frac{1}{3^d} = q$,  which for $d=1$ would be trivially wrong. In
this sense $d$ is a measure of the ``deformation'' of the
$u_{1,q}$-space {\it with respect to the $u_0$-space}. In other
words, we require that the measure of the deformed segment $u_{1,q}$
with respect to the undeformed segment $u_0$ be $1$:
$\frac{u_{1,q}}{u_0} = 1$, namely $\al q = \frac{4}{3^{d}} = 1$. In
the following, for brevity I will thus set $u_0  = 1$, whenever no
misunderstanding arises.

After having partitioned $u_0$ in three equal segments, since the
deformation of $u_0$ into $u_{1,q}$ is performed by varying the
number $\al$ of such segments from $3$ to $4$, we expect that $\al$
and its derivative $\frac{d}{d\al}$ play a relevant role in the
fractal structure. We will see indeed that $(\al, \frac{d}{d\al})$
play the role of conjugate variables (cf. Eq. (\ref{(aaex2.7)}).


Steps of higher order $n$, $n = 2,3,4,..\infty$, can be obtained by
iteration of the deformation process keeping $q = \frac{1}{3^d}$ and
$\alpha = 4$ . For example, in the step $n =2$, we rescale $L_1$ by
a further factor $3$, $L_2 = \frac{1}{3} L_1$, and construct the
deformed segment $u_{2,q}(\alpha) \equiv q \, \alpha \,
u_{1,q}(\alpha) = (q \, \alpha)^{2} \, u_0$, and so on. For the
$n$th order deformation we have
\be \lab{31a} u_{n,q}(\alpha) \equiv (q \, \alpha) \,
u_{n-1,q}(\alpha) ~ , \quad n = 1,2,3,... \ee
i.e., for any $n$
\be \lab{31} u_{n,q}(\alpha)  = (q \, \alpha)^{n} \, u_0 ~ . \ee

By proceeding by iteration, 
or, equivalently, by requiring that $\frac{u_{n,q}(\alpha)}{u_0}$ be
$1$ for any $n$, gives $(q \, \alpha)^{n} = 1$ and Eq. (\ref{30}) is
again obtained.  Notice that the fractal is mathematically defined
in the limit of infinite iterations of the deformation process, $n
\rar \infty$: in this sense, the fractal is the limit of the
deformation process for $n \rar \infty$. As a matter of facts, the
definition of fractal dimension is  given starting from $(q \al)^n
=1$ in the $n \rar \infty$ limit. Since $L_n \rar 0$ for $n \rar
\infty$, the Koch fractal is a curve which is everywhere
non-differentiable \cite{Peitgen}.

Provided one specializes the values of the ``deformation'' parameter
$q$ and of the $\alpha$ parameter, Eqs. (\ref{31a}) and (\ref{31})
fully characterize the fractal. They express in analytic form, {\it
in the $n \rar \infty$ limit}, the {\it self-similarity} property of
a large class of fractals (the Sierpinski gasket and carpet, the
Cantor set, etc.): ``cutting a piece of a fractal and magnifying it
isotropically to the size of the original, both the original and the
magnification look the same'' \cite{Bunde}. In this sense one also
says that fractals are ``scale free'', namely viewing a picture of
part of a fractal one cannot deduce its actual size if the unit of
measure is not given in the same picture \cite{BakinBunde}. It has
to be stressed that only in the $n \rar \infty$ limit
self-similarity is defined (self-similarity does not hold when
considering only a finite number $n$ of iterations).  I also recall
that invariance (always in the limit of $n \rar \infty$ iterations)
only under anisotropic magnification is called self-affinity. The
discussion below can be extended to self-affine fractals. I will not
discuss here the measure of lengths in fractals, the Hausdorff
measure, the fractal ``mass'' and other fractal properties. The
reader is referred to the existing literature.

My main observation is now that Eq. (\ref{31}) expresses the deep
formal connection with the theory of coherent states and the related
algebraic structure.  I discuss this in the next subsection.

\subsection{Fractals and deformed coherent states}\label{secVit.51}

In order to make the connection between fractals and the algebra of
coherent states explicit, I observe that, by considering in full
generality the complex $\al$-plane, the functions
\be \lab{(a2.8)} u_n(\al) = {\al^n\over \sqrt{n!}} ~,~\quad u_0
(\al) = 1 ~~ \quad \quad~~ n\in \mathcal{N}_+ ~, \quad \al \in
{\cal C} ~,
\ee
form in the space ${\cal F}$ of the entire analytic functions a
basis which is orthonormal under the gaussian measure $d\mu(\al)=$
$\displaystyle{{1\over{\pi}} e^{- |\al|^2} d\al d{\bar \al}}$. In
Eq. (\ref{(a2.8)}) the factor $\frac{1}{\sqrt{n!}}$ has been
introduced to ensure the normalization condition with respect to the
gaussian measure.

The functions $u_{n,q}(\alpha)|_{q \rar 1}$ in Eq. (\ref{31}) (for
the factor $q \neq 1$ see the discussion below), are thus
immediately recognized to be nothing but the restriction to real
$\al$ of the functions in Eq. (\ref{(a2.8)}), apart the
normalization factor $\frac{1}{\sqrt{n!}}$. The study of the fractal
properties may be thus carried on in the space ${\cal F}$ of the
entire analytic functions, by restricting, at the end, the
conclusions to real $\al$, $\al \rar {\it Re}(\al)$. Furthermore,
since actually in Eq. (\ref{31}) it is $q \neq 1$ ($q < 1$), one
also needs to consider the ``$q$-deformed'' algebraic structure of
which the space ${\cal F}$ provides a representation. This is indeed
the plan I will follow below.

Let me start by observing that the space ${\cal F}$  is a vector
space which provides the so called Fock-Bargmann representation
(FBR) \cite{Perelomov:1986tf} of the Weyl--Heisenberg algebra
generated by the set of operators $\{ a, a^{\dagger}, {\bf 1} \}$:
\be \lab{(aex2.7)} [a, \,a^\dagger] = {\bf 1}~, \qquad [N,
\,a^\dagger] = a^\dagger~, \qquad [N, \,a] = - a ~, \ee
where $N \equiv a^{\dagger} a$, with the identification:
\be \lab{(aaex2.7)} N \to \al {d\over d\al} ~ ,~\quad a^\dagger \to
\al ~ ,~\quad a \to
{d\over d \al} ~ . 
\ee
%

The $u_n(\al)$ (Eq. (\ref{(a2.8)})) are easily seen to be eigenkets
of $N$ with integer (positive and zero) eigenvalues. The FBR is the
Hilbert space ${\cal K}$ generated by the $u_n(\al)$, i.e. the whole
space ${\cal F}$ of entire analytic functions. Any vector
$\displaystyle{|\psi \rangle}$ in ${\cal K}$ is associated, in a
one-to-one correspondence, with a function $\psi (\al) \in {\cal F}$
and is thus described by the set $\{c_n ; ~c_n\in \mathcal{C},
~\sum_{n=0}^\infty |c_n|^2 = 1 \}$ defined by its expansion in the
complete orthonormal set of eigenkets $\{ |n\rangle \}$ of $N$:
\bea \lab{(aex2.9)} |\psi \rangle  = \sum_{n=0}^\infty c_n |n\rangle
&\rightarrow&  \psi (\al) = \sum_{n=0}^\infty c_n u_n(\al), \\
\lab{(aex2.9c)} \langle \psi|\psi \rangle =  \sum_{n=0}^\infty
|c_n|^2 &=& \int |\psi (\al)|^{2} d \mu (\al) = ||\psi||^{2}=1, \\
\lab{(aex2.9b)} |n\rangle  &=& \frac{1}{\sqrt{n!}} (a^\dag)^{n}| 0
\rangle ~, \eea
where $|0\rangle$ denotes the vacuum vector, $a |0\rangle = 0$,
$\langle 0|0 \rangle = 1$. The series expressing $\psi (\al)$ in Eq.
(\ref{(aex2.9)}) converges uniformly in any compact domain of the
$\al$-plane due to the condition $\sum_{n=0}^\infty |c_n|^2 = 1$
(cf. Eq. (\ref{(aex2.9c)})), confirming that $\psi (\al)$ is an
entire analytic function. Note that, as expected in view of the
correspondence ~${\cal K} \to {\cal F}$~ ($~|n\rangle \to
u_n(\al)$),
\be \lab{(aex2.10)} a^\dagger ~u_n (\al) = \sqrt{n + 1} ~u_{n+1}
(\al) ~, ~~\quad a ~u_n (\al) = \sqrt{n} ~u_{n-1} (\al)~,
\ee
\be \lab{(aexg2.11)} N ~u_n (\al) = a^\dagger a ~u_n (\al) =
\al{d\over d \al} ~u_n (\al) = n~ u_n (\al)~.
\ee
Equations (\ref{(aex2.10)}) and (\ref{(aexg2.11)}) establish the
mutual conjugation of $a$ and $a^\dagger$ in the FBR, with respect
to the measure $d\mu (z)$.

%


Note that, upon introducing $H \equiv N + {1\over 2}$, the three
operators $\{ a, a^{\dagger}, H \}$ close, on ${\cal K}$, the
relations
\be \lab{(g2.2)} \{ a, a^\dagger \} = 2 H ~ ,~\quad [ H, a ] = - a ~
,~\quad [ H, a^\dagger ] = a^\dagger ~ ,  \ee
that are equivalent to (\ref{(aex2.7)}) on ${\cal K}$ and  show the
intrinsic nature of superalgebra of such scheme.

The Fock--Bargmann representation provides a simple frame to
describe the usual coherent states (CS)
\cite{Perelomov:1986tf,Klauder} $|\alpha \rangle$:
\be \lab{(3.1)} |\alpha\rangle = {\cal D}(\alpha ) |0\rangle ~
,\quad a |\alpha\rangle = \alpha |\alpha\rangle~ ,  \quad \alpha \in
{\cal C} ~, \ee
\be \lab{(3.2)} |\alpha\rangle = \exp\biggl(-{|\alpha|^2\over 2}
\biggr) \sum_{n=0}^\infty {{\alpha ^n}\over {\sqrt{n!}}} |n\rangle =
\exp\biggl(-{{|\alpha|^2}\over 2}\biggr) \sum_{n=0}^\infty
u_n(\alpha) |n\rangle .  \ee
The unitary displacement operator ${\cal D}(\alpha)$ in
(\ref{(3.1)}) (mentioned in Section \ref{secVit.2}) is given by:
\be \lab{(3.3)} {\cal D}(\alpha) = \exp\bigl(\alpha a^\dagger -{\bar
\alpha} a \bigr) = \exp\biggl(-\frac{|\alpha|^2}{2}\biggr)
\exp\bigl(\alpha a^\dagger\bigr) \exp\bigl(-{\bar \alpha} ~a\bigr)
~. \ee

The explicit relation between the CS and the entire analytic
function basis ~$\{ u_n(\al) \}$ (Eq. (\ref{(a2.8)})) is:
\be \lab{(3.2a)} u_n (\alpha) = {\rm e}^{{1\over 2}|\alpha|^2}
\langle n|\alpha \rangle~. \ee

The following relations hold
\be \lab{(3.4a)} {\cal D}^{-1}(\alpha)~ a ~{\cal D}(\alpha) = a +
\al ~,\ee
\bea \lab{(3.4)} {\cal D}(\alpha) {\cal D}(\beta) &=& \exp\bigl( i
{\it Im}(\alpha {\bar \beta})\bigr) {\cal D}(\alpha + \beta)~ , \\
\lab{(3.5)} {\cal D}(\alpha) {\cal D}(\beta) &=&\exp\bigl( 2 i {\it
Im}(\alpha {\bar \beta})\bigr) {\cal D}(\beta) {\cal D}(\alpha) ~.
\eea

The operator ${\cal D}(\alpha)$ is a bounded operator defined on the
whole ${\cal K}$. Eq. (\ref{(3.4)}) is nothing but a representation
of the Weyl--Heisenberg group usually denoted by $W_{1}$
\cite{Perelomov:1986tf}. It must be remarked that the set $\{
|\alpha \rangle \}$ is an overcomplete set of states, from which,
however, a complete set can be extracted. Is well known that in
order to extract a complete set of CS from the overcomplete set it
is necessary to introduce in the $\alpha$-complex plane a regular
lattice $L$, called the von Neumann lattice
\cite{Perelomov:1986tf,Fock}. For shortness, I will not discuss this
point here, see \cite{Perelomov:1986tf} for a general discussion and
original references (see also \cite{CeleghDeMart:1995} where the von
Neumann lattice is discussed also in connection with the deformation
of the Weyl-Heisenberg algebra introduced below).

Let me now introduce the finite difference operator ${\cal D}_q$,
also called the $q$-derivative operator \cite{13}, defined by:
\be \lab{(2.12)} {\cal D}_q f(\al) = {{f(q \al) - f(\al)}\over
{(q-1) \al}} ~, \ee
with ~$f(\al) \in {\cal F}\; ,\; q = e^\zeta \; ,\; \zeta \in
{
{\cal C}}$ . ${\cal D}_q$ reduces to the standard derivative for $q
\to 1$ ($\zeta \to 0$).
%
In the space ${\cal F}$, ${\cal D}_q$ satisfies, together with $\al$
and $\al {d\over {d\al}}$, the commutation relations:
\be \lab{(2.17)} \bigl[ {\cal D}_q , \al \bigr] = q^{\al {d\over
{d\al}}} \quad ,\quad \left [ \al {d\over d\al} , {\cal D}_q \right
] = - {\cal D}_q \quad ,\quad \left [\al {d\over d\al} , \al \right
] = \al ~, \ee
which, as for Eq. (\ref{(aaex2.7)}), lead us to the  identification
\be \lab{(2.18)} N \to \al {d\over d\al} \quad ,\quad {\hat a}_q \to
\al \quad ,\quad a_q \to {\cal D}_q ~, \ee
with ~${\hat a}_q = {\hat a}_{q=1} = a^\dagger$~ and ~$\lim_{q\to 1}
a_q = a$ on ${\cal F}$. With such an identification the algebra
(\ref{(2.17)}) can be seen as the $q$-deformation of the algebra
(\ref{(aex2.7)}).
For shortness I omit to discuss further the properties of ${\cal
D}_q $ and the $q$-deformed algebra (\ref{(2.17)}). More details can
be found in Ref. \cite{CeleghDeMart:1995}. Here I only recall that
the relations analogous to (\ref{(aex2.10)}) for the $q$-deformed
case are
\be \lab{(2.19)} {\hat a}_q u_n (\al) = \sqrt{n + 1} ~u_{n+1}
(\al)~, \quad a_q u_n (\al) = q^{{n-1}\over {2}} {{[n]_q}\over
{\sqrt{n}}} ~u_{n-1} (\al)~, \ee
where $[n]_q \equiv  {{q^{{1\over 2}n}-q^{-{1\over
2}n}}\over{q^{1\over 2} - q^{-{1\over 2}}}}$~, and that the operator
$q^N$ acts on the whole ${\cal F}$ as
\be \lab{(2.20)}  q^{N} f(\al) = f(q \al) ~, \quad f(\al) \in {\cal
F}~. \ee
This result was originally obtained in Ref.
\cite{CeleghDeMart:1995}. There it has been remarked that since the
$q$-deformation of the algebra has been essentially obtained by
replacing the customary derivative with the finite difference
operator, then a deformation of the operator algebra acting in
${\cal F}$ should arise whenever one deals with some finite scale,
e.g. with some discrete structure, lattice or periodic system
(periodicity is but a special invariance under finite difference
operators), which cannot be reduced to the continuum by a limiting
procedure. The $q$-deformation parameter is related with the finite
scale. A finite scale occurs indeed also in the present case of
fractals and therefore also in this case we expect and in fact have
a deformation of the algebra.

Eq. (\ref{(2.20)}) applied to the coherent state functional
(\ref{(3.2)}) gives
\be \lab{(a2.21)}  q^{N} |\al \rangle = |q \al \rangle =
\exp\biggl(-{{|q\alpha|^2}\over 2}\biggr) \sum_{n=0}^\infty \frac{(q
\alpha)^{n}}{\sqrt{n!}}~ |n\rangle~, \ee
and, since $q \al \in 
{\cal C}$, from Eq. (\ref{(3.1)}),
\be \lab{(ab2.21)}  a~ |q \al \rangle = q \al ~|q \al \rangle ~,
\quad q \al \in 
{\cal C}~. \ee
By recalling that we have set $u_0 \equiv 1$, the $n$th fractal
iteration, Eq. (\ref{31}), is obtained by projecting out the $n$th
component of $|q \al \rangle$ and restricting to real $q \al$, $q
\al \rar {\it Re} (q \al)$:
\be \lab{(a2.22)} u_{n,q} (\al) = (q \alpha)^{n} = {\sqrt{n!}}~
\exp\biggl({{|q\alpha|^2}\over 2}\biggr) \langle n|q \al \rangle,
\quad {\rm for~ any}~ n, ~~ q \al \rar {\it Re} (q \al).\ee
Taking into account that $\langle n| = \langle 0|
~\frac{(a)^{n}}{\sqrt{n!}}$,  Eq. (\ref{(a2.22)}) gives
\be \lab{(a2.23)} u_{n,q} (\al) = (q \alpha)^{n} =
\exp\biggl({{|q\alpha|^2}\over 2}\biggr) \langle 0|(a)^{n}|q \al
\rangle, \quad {\rm for~ any}~ n, ~~ q \al \rar {\it Re} (q \al),
\ee
which shows that the operator $(a)^{n}$ acts as a ``magnifying''
lens \cite{Bunde}: the $n$th iteration of the fractal can be
``seen'' by applying $(a)^{n}$ to $|q \al \rangle$ and restricting
to real $q \al$:
\be \lab{(ab2.24)}  \langle q \al | (a)^{n} |q \al \rangle =  (q
\alpha)^{n} = u_{n,q} (\al), \quad q \al \rar {\it Re} (q \al). \ee
Note that the equivalence between Eq. (\ref{(a2.23)}) and Eq.
(\ref{(ab2.24)}) stems from the coherent state properties
\cite{Perelomov:1986tf}.

Eq. (\ref{(ab2.21)})  expresses the invariance of the coherent state
representing the fractal under the action of the operator
$\frac{1}{q \al}a$~. This reminds us of the fixed point equation
$W(A) = A$, where $W$ is the Hutchinson operator \cite{Bunde},
characterizing the iteration process for the fractal $A$ in the $n
\rar \infty$ limit. Such an invariance property allows to consider
the coherent functional $\psi (q\al)$ as an ``attractor'' in
$
{\cal C}$.

In conclusion, the meaning of Eq. (\ref{(a2.21)}) is that the
operator $q^{N}$ applied to $|\al \rangle$ ``produces'' the fractal
in the functional form of the coherent state $|q \al \rangle$. The
$n$th fractal stage of iteration, $n = 0,1,2,..,\infty$ is
represented, in a one-to-one correspondence, by the $n$th term in
the coherent state series in Eq. (\ref{(a2.21)}). I call $q^{N}$
{\it the fractal operator}.

Eqs. (\ref{(a2.22)}), (\ref{(a2.23)}) and (\ref{(ab2.24)}) formally
establish the searched connection between fractals and the
($q$-deformed) algebra of the coherent states.

\subsection{Fractals, squeezed coherent states and
noncommutative geometry in the phase space}\label{secVit.52}

Let me now look at the fractal operator $q^N$ from a different
perspective. I consider the identity
\be \lab{(4.1)} 2 \al {d\over {d\al}} \psi (\al) = \left\{{1\over 2}
\left[\left(\al + {d\over {d\al}}\right)^2 - \left(\al - {d\over
{d\al}}\right)^2\right] - 1\right\} \psi (\al) ~ , \ee
which holds in the Hilbert space identified with the space ~${\cal
F}$ of entire analytic functions $\psi (\al)$. It is convenient to
set $\al \equiv x + iy$, $x$ and $y$ denoting the real and the
imaginary part of $\al$, respectively. I then introduce the
operators
\be \lab{(4.2a)} {c} = {1\over {\sqrt{2}}} \bigl( \al +  {d\over
{d\al}} \bigr) \quad ,\quad { c}^\dagger = {1\over {\sqrt{2}}}
\bigl( \al - {d\over {d\al}}\bigr)\quad , \quad [c, c^\dagger ] =
{\bf 1} ~. \ee
Their relation with the FBR operators $a$ and $a^\dagger$ is
\be \lab{(4.2b)} \al = {1\over {\sqrt{2}}} ( c + {c}^\dagger )~ \to
a^\dagger ~ , \quad {d\over {d\al}} = {1\over {\sqrt{2}}} ( c -
{c}^\dagger ) ~\to a ~.  \ee

In ~${\cal F}$, $~{c}^\dagger$ is indeed the conjugate of $c$
\cite{Perelomov:1986tf,CeleghDeMart:1995}. In the limit $\al \rar
{\it Re} (\al)$, i.e.  $y \to 0$,~ $c$ and $c^\dagger$ turn into the
conventional annihilation and creator operators associated with $x$
and $p_x$ in the canonical configuration representation,
respectively. I now remark that the fractal operator $q^N$ can be
realized in  ~${\cal F}$ as:
\be \lab{(4.3)} q^N \psi (\al) = {1\over{\sqrt q}}~\exp\biggl({
\zeta \over 2}\bigl(c^2 - {c^\dagger}^2\bigr)\biggr)\psi (\al)
\equiv {1\over{\sqrt q}} {\hat {\cal S}(\zeta)} \psi (\al) \equiv
{1\over{\sqrt q}} \psi_{s}(\al) ~,  \ee
where $q = e^{\zeta}$ (for simplicity, assumed to be real) and as
usual $N = \al {d\over {d \al}}$. From Eq. (\ref{(4.3)}) we see that
$q^N$ acts in ~${\cal F}$, as well as in the configuration
representation in the limit $y \to 0$, as the squeezing operator
${\hat {\cal S}(\zeta)}$ (well known in quantum optics
\cite{CeleghDeMart:1995,Yuen:1976vy}) up to the numerical factor
${1\over{\sqrt q}}$. $\zeta = \ln q $ is called the squeezing
parameter. In (\ref{(4.3)}) $\psi _s(\al)$ denotes the squeezed
states in FBR.

Since $ q^N \psi (\al) =  \psi (q \al)$ (cf. Eq. (\ref{(2.20)})),
from Eq. (\ref{(4.3)}) we see that the $q$-deformation process,
which we have seen is associated to the fractal generation process,
is equivalent to the squeezing transformation.

Note that the right hand side of (\ref{(4.3)}) is an $SU(1,1)$ group
element. In fact, by defining $K_{-} = {1\over 2}c ^2$, $K_{+} =
{1\over 2}c^{\dagger 2}$, $K_{\al} = {1\over 2}(c^\dagger c +
{1\over 2})$, one easily checks they close the algebra $su(1,1)$. We
indeed obtain the $SU(1,1)$ (Bogoliubov) transformations for the
$c$'s operators:
\bea \lab{(4.4b)} {\hat {\cal S}}^{-1}(\zeta) ~c~ {\hat {\cal
S}}(\zeta) = c ~\cosh \zeta - {c}^\dagger ~\sinh \zeta~,\\
{\hat {\cal S}}^{-1}(\zeta) ~{c}^\dagger ~{\hat {\cal S}}(\zeta) =
{c}^\dagger ~\cosh \zeta -  c ~\sinh \zeta ~, \eea
and in the $y \to 0$ limit (still in ${\cal F}$)
\bea \lab{(4.4c)}  {\hat {\cal S}}^{-1}(\zeta) ~c~ {\hat {\cal
S}}(\zeta) &\rar& {\hat {\cal S}}^{-1}(\zeta) ~a~ {\hat {\cal
S}}(\zeta)
= a ~\cosh \zeta - {a}^\dagger ~\sinh \zeta\\
\lab{(4.4c)} {\hat {\cal S}}^{-1}(\zeta) ~{c}^\dagger ~{\hat {\cal
S}}(\zeta) &\rar& {\hat {\cal S}}^{-1}(\zeta)~ a^\dagger ~{\hat
{\cal S}}(\zeta) = {a}^\dagger ~\cosh \zeta -  a ~\sinh \zeta ~.
\eea
%
Moreover, in the $y \to 0$ limit,
\bea \lab{(4.4d)} {\hat {\cal S}}^{-1}(\zeta)~ \al~ {\hat {\cal
S}}(\zeta) &=& {1\over{q}}\al \rar {1\over{q}}x ~ , \\
\lab{(4.4e)}{\hat {\cal S}}^{-1}(\zeta)~p_{\al} ~{\hat {\cal
S}}(\zeta) &=& q p_{\al} \to qp_x ~ ,  \eea
where ${p_\al \equiv -i{d\over{d\al}}}$, and
\bea \lab{(4.5)} \int d\mu(\al) {\bar \psi}(\al){\hat {\cal S}}^{-
1}(\zeta)~ \al ~{\hat {\cal S}} (\zeta) \psi (\al)   &\rar&
{1\over{q}}<x>\\
 \int d\mu(\al) {\bar \psi}(\al){\hat {\cal S}}^{- 1}(\zeta)~
 p_{\al}~
{\hat {\cal S}} (\zeta) \psi (\al)   &\rar&  q<p_{x}>  \eea
so that the root mean square deviations $\Delta x$ and $\Delta
p_{x}$ satisfy
\be \lab{(4.6)} \Delta x \Delta p_{x} = {1\over 2} \quad ,\quad
\Delta{x} = {1\over q} {\sqrt{1\over 2}}\quad ,\quad \Delta {p_{x}}
= q{\sqrt{1\over 2}}~ . \ee
This confirms that the $q$-deformation plays the role of squeezing
transformation. Note that the action variable $\int p_{x}~dx$ is
invariant under the squeezing transformation.

Eq. (\ref{(4.4d)}) shows that $\al \rar {1\over{q}}\al$ under
squeezing transformation, which, in view of the fact that $q^{-1} =
\al$ (cf. Eq. (\ref{30a})), means that $\al \rar \al^{2}$, i.e.
under squeezing we proceed further in the fractal iteration process.
Thus, the fractal iteration process can be described in terms of the
coherent state squeezing transformation.

I recall that by means of the squeezing transformations the
Weyl-Heisenberg representations are labeled by the $q$-parameter and
that in the infinite volume limit (infinite degrees of freedom) they
are unitarily inequivalent representations: different values of the
$q$-deformation parameter label ``different'', i.e. unitarily
inequivalent representations in QFT
\cite{Iorio:1994jk,CeleghDeMart:1995}. By changing the value of the
$q$-parameter one thus moves from a given representation to another
one, unitarily inequivalent to the former one. Besides the scale
parameter one might also consider, phase parameters and translation
parameters characterizing (generalized) coherent states (such as
$SU(2)$, $SU(1,1)$, etc. coherent states). For example, by changing
the parameters in a {\it deterministic iterated function process},
also referred to as {\it multiple reproduction copy machine}
process, (such as phases, translations, etc.) the Koch curve may be
transformed into another fractal (e.g. into Barnsley's fern
\cite{Peitgen}). In the scheme here presented, these fractals are
then described by corresponding unitarily inequivalent
representations in the limit of infinitely many degrees of freedom
(infinite volume limit). The trajectories induced by the changes of
the parameters over the space of the representations can be shown to
be, under quite general conditions, chaotic trajectories
\cite{Vitiello:2003me}. This might shed some light on the richness
of the variety of ``different'' fractal shapes obtainable by
changing the parameters of the fractal one starts with
\cite{Peitgen}. Work is in progress on such a subject.

Due to the holomorphy conditions holding for $f(\al)\in {\cal F}$
\be \lab{(4.7)} {d\over d\al} f(\al) = {d\over dx} f(\al) = -i
{d\over dy} f(\al) ~, \ee
in the $y \to 0$ limit we get form (\ref{(4.2a)})
\be \lab{(4.8)} {c} \rar {1\over {\sqrt{2}}} \bigl( x + i p_{x}
 \bigr) \equiv {\hat z} ~ ,\quad { c}^\dagger \rar {1\over
{\sqrt{2}}} \bigl( x - i p_{x} \bigr) \equiv {\hat z}^\dagger ~ ,
\quad [{\hat z}, {\hat z}^\dagger ] = 1
 ~, \ee
where $p_{x}= -i {d\over dx}$. ${\hat z}^\dagger$ and  ${\hat z}$
are the usual creation and annihilation operators  in the
configuration representation.  Under the action of the squeezing
transformation, use of (\ref{(4.4d)}) and (\ref{(4.4e)}) leads to
\be \lab{(4.8)} {\hat z}_{q} = {1\over {q \sqrt{2}}} \bigl( x + i
q^{2} p_{x}
 \bigr)  ~ ,\quad {\hat z}^\dagger_{q} = {1\over
{q \sqrt{2}}} \bigl( x - i q^{2} p_{x} \bigr) , \quad [{\hat
z}_{q},{\hat z}^\dagger_{q} ] = 1 
 ~. \ee
Notice that the ${\hat z}$ and ${\hat z}^\dagger$ algebra is
preserved under the squeezing transformation, which  is indeed a
canonical transformation. In the ``deformed'' phase space let us
denote the coordinates $(x,q^{2}p_{x})$ as $(x_{1},x_{2})$, where
$x_{1} \equiv x$ and $x_{2} \equiv q^{2}p_{x}$. Coordinates do not
commute in this (deformed) space:
\be \lab{(4.8)}  [x_{1},x_{2}] = iq^{2} ~. \ee
We thus recognize that $q$-deformation introduces non-commutative
geometry in the $(x_{1},x_{2})$-space. In such a space the
noncommutative Pythagora's theorem gives the distance $D$:
\be \lab{(4.9)}  D^{2} = x_{1}^{2} + x_{2}^{2} = 2 q^{2} \bigl({\hat
z}^\dagger_{q} {\hat z}_{q} + \frac{1}{2} \bigr)~. \ee
In ${\cal F}$,   in the $y \to 0$ limit, form the known properties
of creation and annihilation operators  we then get
\be \lab{(4.10)}  D^{2}_{n} =  2q^{2} \bigl(n + \frac{1}{2} \bigr)~,
\quad n = 0,1,2,3... ~,\ee
i.e. in the space $(x_{1},x_{2})$ associated to the coherent state
fractal representation, the $(x_{1},x_{2})$-distance is quantized
according to the unit scale set by $q$. Eq. (\ref{(4.10)}) also
shows that in the space $(x_{1},x_{2})$ we have quantized ``disks''
of squared radius vector $D^{2}_{n}$. It is interesting to observe
that the ``smallest'' of such disks has non-zero radius given by the
deformation parameter $q$ (recall that $q = \frac{1}{3^d}$ when Koch
fractal is considered). Recalling the expression of the energy
spectrum of the harmonic oscillator, one could write Eq.
(\ref{(4.10)}) as $D^{2}_{n} =  2q^{2} \bigl(n + \frac{1}{2} \bigr)
\equiv E_{n}$, $n = 0,1,2,3...$, where $E_{n}$ might be thought as
the ``energy'' associated to the fractal $n$-stage.

\section{Final remarks}\label{secVit.6}

Let me close the paper by observing that in the case of
topologically non-trivial condensation at finite temperature the
order parameter $v(x, \beta)$ provides a mapping between the domains
of variation of $(x, \beta)$ and the space of the unitarily
inequivalent representations of the canonical commutation relations.
As well known, this is the homotopy mapping between the $(x,\beta)$
variability domain and the group manifold. In the vortex case one
has the mapping $\pi$ of $S^{1}$, surrounding the $r = 0$
singularity, to the group manifold of U(1) which is topologically
characterized by the winding number $n \in Z \in \pi_{1}(S^{1})$.
Such a singularity at $r = 0$ is carried by the boson condensation
function of the NG modes.  In the monopole case \cite{Manka:1988tb},
the mapping $\pi$ is the one of the sphere $S^{2}$, surrounding the
singularity $r=0$, to SO(3)/SO(2) group manifold, with homotopy
classes of $\pi_{2}(S^{2})= Z$. Same situation occurs in the
sphaleron case \cite{Manka:1988tb}, provided one replaces SO(3) and
SO(2) with SU(2) and U(1), respectively.

As discussed in the previous sections, transitions between phases
characterized by an order parameter imply "moving" over unitarily
inequivalent representations, and this is achieved by gradients in
NG boson condensation function. In the presence of a gauge field,
macroscopic ground state field and currents can only be obtained by
non-homogeneous NG boson condensation with topological
singularities. The occurrence of such topologically non-trivial
condensation allows the formation of topological defects. This
explains why topological defect formation is observed in symmetry
breaking phase transition processes.

Finite volume effects, effective mass of the NG bosons and
temperature effects have been briefly discussed and related to the
ordered domain size. Correlation modes with non-vanishing effective
mass generate domains which tend to break down into smaller, more
stable domains. An interesting development which can be pursued in a
future research is the one referring to the stability of macroscopic
correlated domains occurring in biology, finance, etc., to which our
conclusions might extend by exploiting the remarkable interplay
shown to emerge between the microscopic dynamics of the system
components and the macroscopic features of the system. As I have
observed, the analysis presented in the previous sections might
suggest, for example, that, within proper conditions, the natural
evolution of the global market is its breakdown into separate parts,
unless, of course, an (enormous) amount of energies is pumped in.

Finally, I have discussed the functional realization of fractal
properties, with particular reference to self-similarity property,
in terms of the $q$-deformed algebra of coherent states. The
relation of fractals with the squeezing operator and noncommutative
geometry in phase space has been also exhibited. Fractal study can
be thus incorporated in the theory of entire analytical functions.

On the other hand, the discussion presented above also shows that
the reverse is also true: under convenient choice of the
$q$-deformation parameter and by a suitable restriction to real
$\al$, coherent states exhibit fractal properties in the
$q$-deformed space of the entire analytical functions.

Since both, fractal structures and coherent states are recognized to
appear in an enormous number of systems and natural phenomena, the
above conclusions may be of large interest in many applications.
Moreover, the relation here established between fractals and
coherent states introduces dynamical considerations in the study of
fractals and of their origin, as well as geometrical insight in the
coherent states properties.

I also remark that fractals are global systems arising from local
deformation processes. Therefore they cannot be purely geometric
objects. Their deep connection with coherent states is therefore not
only expected, but, I would say, necessary, since coherence is the
only available tool existing in our knowledge of physical phenomena
able to provide long range (macroscopic) correlations out of the
microscopic dynamics of elementary components.

In this paper I have not considered ``random fractals'', i.e. those
fractals obtained by randomization processes introduced in their
iterative generation. Their characteristics suggest that they must
be related with dissipative systems and since self-similarity is
still a characterizing property in many of such random fractals, my
conjecture is that also in such cases there must exist a deep
connection with the coherent state algebraic structure. This will be
the subject of future work.


\section{Acknowledgements}


This work has been partially supported by INFN and Miur.


\end{document}